\documentclass[11pt]{article}

\makeatletter
\@addtoreset{equation}{section}
    
\makeatother

\usepackage{amsmath,amssymb}

\begin{document}
%%%%%%%%%%%%%%%%

\title{
{\bf Majorana equation and exotics: higher derivative models, anyons
and noncommutative geometry}}

\author
{{\sf Mikhail S. Plyushchay}\thanks{E-mail: mplyushc@lauca.usach.cl}
\\[4pt]
 {\small \it Departamento de F\'{\i}sica,
Universidad de Santiago de Chile}\\
{\small \it Casilla 307, Santiago 2, Chile}\\
}
\date{}

\maketitle
%\null\vskip-15mm

\begin{abstract}
In 1932 Ettore Majorana proposed  an infinite-component relativistic
wave equation for  particles of arbitrary integer and half-integer
spin. In the late 80s and early 90s it was found that the
higher-derivative geometric particle models underlie the Majorana
equation, and that its  (2+1)-dimensional analogue provides with a
natural basis for the description of relativistic anyons. We review
these aspects  and discuss the  relationship of the equation to the
exotic planar Galilei symmetry and noncommutative geometry. We also
point out the relation of some Abelian gauge field theories with
Chern-Simons terms to the Landau problem in the noncommutative plane
from the perspective of the Majorana equation.

\end{abstract}

%\vskip 0.5cm

\vskip.5cm\noindent

\section{Introduction}

Ettore Majorana was the first to study the infinite-component
relativistic fields. In the pioneering 1932 paper \cite{Maj}, on the
basis of the linear differential wave equation of a Dirac form, he
constructed  a relativistically invariant theory for arbitrary
integer or half-integer spin particles. It was the first
recognition, development and application of  the
infinite-dimensional unitary representations of the Lorentz group.
During a long period of time, however, the Majorana results remained
practically unknown, and the theory was rediscovered in 1948 by
Gel'fand and Yaglom \cite{GelYa} in a more general framework of the
group theory representations. In 1966 Fradkin revived the Majorana
remarkable work (on the suggestion of Amaldi) by translating it into
English and placing it in the context of the later research
\cite{Frad}. In a few years the development of the concept of the
infinite-component fields \cite{Fro1}--\cite{BoLoToOk} culminated in
the construction of the dual resonance models and the origin of the
superstring theory \cite{Venez}--\cite{GSW}.

After the revival, the Majorana work inspired an interesting line of
research based on a peculiar property of his equation: its time-like
solutions describe \emph{positive energy } states lying on a Regge
type trajectory, but with unusual  dependence of the mass, $M$, on
the spin, $s$, $M_s \propto (const + s)^{-1}$. In 1970, Dirac
\cite{Dirac}  proposed a  covariant \emph{spinor set} of linear
differential equations for the infinite-component field, from which
the Majorana and Klein-Gordon equations appear in the form of
integrability (consistency) conditions. As a result, the new Dirac
relativistic equation describes a massive, spin-zero positive-energy
particle. Though this line of research \cite{Stau1}--\cite{MukDam}
did not find essential development, in particular, due to the
problems arising under the attempt to introduce electromagnetic
interaction, recently it was pushed \cite{HP1}--\cite{HP3} in the
unexpected direction related to the anyon theory
\cite{LeiMyr}--\cite{PanyDef}, exotic Galilei symmetry
\cite{LL}--\cite{HorMar}, and non-commutative geometry
\cite{SeiWit}--\cite{ConS}.

In pseudoclassical relativistic particle model associated with  the
quantum Dirac spin-1/2 equation, the spin degrees of freedom are
described by the \emph{odd} Grassmann variables \cite{BerMar}. In
1988 it was observed \cite{Pcurv1} that the (3+1)D particle analogue
of the Polyakov string with rigidity \cite{Pol1} possesses the mass
spectrum of the \emph{squared} Majorana equation. The  model of the
particle with rigidity contains, like the string model \cite{Pol1},
the higher derivative curvature term in the action. It is this
higher derivative term that effectively supplies the system with the
\emph{even} spin degrees  of freedom of noncompact nature and leads
to the infinite-dimensional representations of the Lorentz group.
Soon it was found that the quantum theory of another higher
derivative model of the (2+1)D relativistic particle with torsion
\cite{Ptor}, whose Euclidean version underlies the Bose-Fermi
transmutation mechanism \cite{Pol2}, is described by the
\emph{linear} differential infinite-component wave equation of the
Majorana form. Unlike the original Majorana equation, its (2+1)D
analogue provides with the quantum states of any (real) value of the
spin, and so, can serve as a basis for the construction of
relativistic anyon theory \cite{PlAny}--\cite{PanyDef}. It was shown
recently \cite{HP1,HP2} that the application of the special
non-relativistic limit ($c\rightarrow \infty$, $s\rightarrow
\infty$, $s/c^2\rightarrow \kappa=const$) \cite{JNlim,DHNR} to the
model of relativistic particle with torsion produces the higher
derivative model of a planar particle \cite{LSZ} with associated
exotic (two-fold centrally extended) Galilei symmetry \cite{LL}. The
quantum spectrum of the higher derivative model \cite{LSZ}, being
unbounded from below, is described by reducible representations of
the exotic planar Galilei group. On the other hand, the application
of the same limit to the (2+1)D analogue of the Dirac spinor set of
anyon equations \cite{PanyDef} gives rise to the
Majorana-Dirac-Levy-Leblond type infinite-component wave equations
\cite{HP2}, which describe irreducible representations of the exotic
planar Galilei group corresponding to a free particle with
non-commuting coordinates \cite{DH}.

Here we review the described relations of the Majorana equation to
the higher derivative  particle models, exotic Galilei symmetry and
associated noncommutative structure. We also discuss the
relationship of the (2+1)D relativistic Abelian gauge field theories
with Chern-Simons terms \cite{Schon}--\cite{DesJack3} to the Landau
problem in the noncommutative plane \cite{HP3,DH,HorLan,P06} from
the perspective of the Majorana equation.

\section{Majorana equation and Dirac spinor set of equations}

Majorana equation \cite{Maj} is a linear differential equation of
the Dirac form,
\begin{equation}\label{Maj}
    \left( P^\mu \Gamma_\mu-m\right)\Psi(x)=0,
\end{equation}
with $P^\mu=i\partial^\mu$ and
 matrices $\Gamma_\mu$ generating the
Lorentz group via the  anti-de Sitter SO(3,2) commutation relations
similar to those satisfied by the usual
$\gamma$-matrices\footnote{We use the metric with signature
($-,+,+,+$).},
\begin{equation}\label{Gamma}
    [\Gamma_\mu,\Gamma_\nu]=iS_{\mu\nu},\qquad
    [S_{\mu\nu},\Gamma_\lambda]=i(
    \eta_{\nu\lambda}\Gamma_\mu - \eta_{\mu\lambda} \Gamma_\nu),
\end{equation}
\begin{equation}\label{SS}
    [S_{\mu\nu},S_{\lambda\rho}]=i(\eta_{\mu\rho}S_{\nu\lambda} - \eta_{\mu\lambda}S_{\nu\rho}
    +\eta_{\nu\lambda}S_{\mu\rho}-\eta_{\nu\rho}S_{\mu\lambda}).
\end{equation}
The original Majorana realization of the $\Gamma_\mu$ corresponds to
the infinite-dimensional unitary representation of the Lorentz group
in which its Casimir operators $C_1$ and $C_2$ and the Lorentz
scalar $\Gamma_\mu\Gamma^\mu$ take the values
\begin{equation}\label{CCC}
    C_1\equiv
    \frac{1}{2}S_{\mu\nu}S^{\mu\nu}=-\frac{3}{4},\qquad C_2\equiv
    \epsilon^{\mu\nu\lambda\rho}S_{\mu\nu}S_{\lambda\rho}=0,\qquad
    \Gamma_\mu\Gamma^\mu=-\frac{1}{2}.
\end{equation}
A representation space corresponding to (\ref{CCC}) is a direct sum
of the two irreducible SL(2,C) representations characterized by the
integer, $j=0,1,\ldots$,  and half-integer, $j=1/2, 3/2,\ldots$,
values of the SU(2) subalgebra Casimir operator, $M_i^2=j(j+1)$,
$M_i\equiv \frac{1}{2}\epsilon_{ijk}S^{jk}$. In both cases the
Majorana equation (\ref{Maj}) has time-like (massive), space-like
(tachyonic) and light-like (massless) solutions. The spectrum in the
light-like sector is
\begin{equation}\label{Mj}
    M_j=\frac{m}{j+\frac{1}{2}}, \qquad j= s + n,\qquad
    n=0,1,\ldots,\qquad
    s=0\quad {\rm or}\quad \frac{1}{2}.
\end{equation}
The change $\Gamma_\mu\rightarrow -\Gamma_\mu$ in accordance with
(\ref{Gamma}), (\ref{SS}) does not effect on representations of the
Lorentz group as a subgroup of the SO(3,2). For the Majorana choice
with the diagonal generator $\Gamma_0$,
\begin{equation}\label{G0}
        \Gamma_0=j+\frac{1}{2},
\end{equation}
 Eq. (\ref{Maj}) has the time-like ($P^2<0$) solutions
 with positive energy.

In \cite{Dirac},  Dirac suggested an interesting modification of the
Majorana infinite-component theory that effectively singles out the
lowest spin zero time-like state from all the  Majorana equation
spectrum. The key idea  was to generate the Klein-Gordon and
Majorana wave equations via the integrability conditions for some
covariant set of linear differential equations.  Dirac covariant
spinor set of (3+1)D equations has the form
\begin{equation}\label{Dirac}
    {\cal D}_{a}\Psi(x,q)=0,\qquad
    {\cal D}_a=( P^\mu \gamma_\mu + m)_{ab}Q_b,
\end{equation}
where $\gamma$-matrices are taken in the Majorana representation,
and $Q_a=(q_1,q_2,\pi_1,\pi_2)$ is composed from the mutually
commuting dynamical variables $q_\alpha$, $\alpha=1,2$, and
commuting conjugate momenta $\pi_\alpha$,
$[q_\alpha,\pi_\beta]=i\delta_{\alpha\beta}$, while $\Psi(x,q)$ is a
single-component wave function. The SO(3,2) generators are realized
here as quadratic in $Q$ operators,
$$
\Gamma_\mu=\frac{1}{4}\bar{Q}\gamma_\mu Q, \qquad
S_{\mu\nu}=\frac{i}{8}\bar{Q}[\gamma_\mu,\gamma_\nu] Q,
$$
where $\bar{Q}=Q^t\gamma^0$. The covariance of the set of equations
(\ref{Dirac}) follows from the commutation relations
$[S_{\mu\nu},Q]=-\frac{i}{4}[\gamma_\mu,\gamma_\nu]Q$, which mean
that the $Q_a$ is transformed as a Lorentz spinor, and so, the set
of four equations (\ref{Dirac}) is the spinor set. Note also that
$[\Gamma_\mu,Q]=\frac{1}{2}\gamma_\mu Q$, and the $Q_a$ anticommute
between themselves for a linear combination of the SO(3,2)
generators. This means that the $Q_a$, $\Gamma_\mu$ and $S_{\mu\nu}$
generate a supersymmetric extension of the anti-de Sitter algebra.

The Klein-Gordon,
\begin{equation}\label{KG}
    (P^2+m^2)\Psi=0,
\end{equation}
and the Majorana equations (with the parameter $m$ changed in the
latter for $\frac{1}{2}m$) are the integrability conditions for the
spinor set of equations (\ref{Dirac}) \cite{Dirac}. Taking into
account that the $\Gamma_0=\frac{1}{4}(q_1^2+q_2^2+\pi_1^2+\pi_2^2)$
coincides up to the factor $\frac{1}{2}$ with the Hamiltonian of a
planar isotropic oscillator, one finds that the possible eigenvalues
of the $\Gamma_0$ are given by the sets $j=0,1,\ldots$ and
$j=1/2,3/2,\ldots$ in correspondence with Eq. (\ref{G0}). The former
case corresponds to the $\Gamma_0$ eigenstates given by the even in
$q_\alpha$ wave functions, while the latter case corresponds to the
odd eigenstates. Having in mind the Majorana equation spectrum
(\ref{Mj}) (with the indicated change of the mass parameter)
 and Eq. (\ref{KG}), one concludes that the spinor set of equations
(\ref{Dirac}) describes the positive energy spinless
states\footnote{Staunton \cite{Stau2} proposed a modification of the
Dirac spinor set of equations that describes the spin-$1/2$
representation of the Poincar\'e group} of the fixed mass.

\section{Higher derivative relativistic particle models}

The model of relativistic particle with curvature
\cite{Pcurv1,Pisar,Nester,PKuz},
being an analogue of the model of relativistic string with rigidity
\cite{Pol1}, is given by the reparametrization invariant action
\begin{equation}\label{Acurv}
    A=-\int(m+\alpha k)ds,
\end{equation}
where $ds^2=-dx_\mu dx^\mu$, $\alpha>0$ is a dimensionless
parameter\footnote{For $\alpha<0$ the equations of motion of the
system have the only solutions corresponding to the curvature-free
case $\alpha=0$ of a spinless particle of mass $m$ \cite{Pcurv1}.},
and $k$ is the worldline curvature, $k^2=x''_\mu x''{}^\mu$,
$x'_\mu=dx_\mu/ds$. In a parametrization $x_\mu=x_\mu(\tau)$,
Lagrangian of the system is $L=-\sqrt{-\dot{x}^2}(m+k)$, where we
assume that the particle moves with the velocity less than the speed
of light, $\dot{x}^2<0$, $\dot{x}_\mu=dx_\mu/d\tau$, and then
$k^2=(\dot{x}^2\ddot{x}{}^2-(\dot{x}\ddot{x}{}^2)^2)/(\dot{x}^2)^3\geq
0$ \cite{Pcurv1}.
 The Lagrangian equations of motion
have the form of the conservation law of the energy-momentum vector,
\begin{equation}\label{EnMom}
    \frac{d}{d\tau}P_\mu=0,\qquad
    P_\mu=\frac{\partial L}{\partial
    \dot{x}^\mu}-\frac{d}{d\tau}\left(\frac{\partial L}{\partial
    \ddot{x}{}^\mu}\right).
\end{equation}
The dependence of the Lagrangian on  higher derivatives supplies
effectively the system with additional translation invariant degrees
of freedom described by the velocity $v_\mu\equiv\dot{x}_\mu$ and
conjugate momentum \cite{Pcurv1}. This higher derivative dependence
is responsible for a peculiarity of the system: though the particle
velocity is less than the speed of light, the equations of motion
(\ref{EnMom}) have the time-like ($P^2<0$), the light-like ($P^2=0$)
and the space-like ($P^2>0$) solutions \cite{Pcurv1}, whose explicit
form was given in \cite{Pcurv1,PKuz}. This indicates on a possible
relation of the model (\ref{Acurv})  to the infinite-component field
theory associated with the Majorana equation. Unlike  the Majorana
system, however, the quantum version of the model (\ref{Acurv}) has
the states of integer spin only, which lie on the nonlinear Regge
trajectory of the form very similar to (\ref{Mj}) \cite{Pcurv1},
\begin{equation}\label{Ml}
    M_l=\frac{m}{\sqrt{1+\alpha^{-2}l(l+1)}},\qquad
    l=0,1,\ldots.
\end{equation}
The choice of the laboratory time gauge $\tau=x^0$ separates here
the positive energy time-like solutions.

Before we pass over to the discussion of a relativistic particle
model more closely related to the original (3+1)D Majorana equation
from the viewpoint of the structure of the spectrum, but essentially
different from it in some important properties, it is worth to note
that the higher derivative dependence of the action does not
obligatorily lead to the tachyonic states. In Ref. \cite{Pmas0} the
model given by the action of the form (\ref{Acurv}) with parameter
$m=0$ was suggested. It was shown there that in the case of
$\dot{x}^2<0$, the model is inconsistent (its equations of the
motion have no solutions), but for $\dot{x}^2>0$ the model is
consistent and describes massless states of the arbitrary, but fixed
integer or half-integer helicities $\lambda=\pm j$, whose values are
defined by the quantized parameter $\alpha$, $\alpha^2=j^2$. The
velocity higher than the speed of light in such a model originates
from the Zittervewegung associated with nontrivial helicity. System
(\ref{Acurv}) with $m=0$ possesses additional local symmetry
\cite{Pmas0,Ramos} (action (\ref{Acurv}) in this case has no scale
parameter), and it is such a gauge symmetry that is responsible for
separation of the two physical helicity components from the
infinite-component Majorana type field (cf. the system given by the
Dirac spinor set of equations (\ref{Dirac})). Recently, the interest
to such a higher derivative massless particle system has been
revived \cite{Mourad,EdMar} in the context of the massless higher
spin field theories \cite{HSF,HSV}.

The (2+1)D relativistic model of the particle with torsion
\cite{Ptor} is given by the action
\begin{equation}\label{Ator}
    A=-\int(m+\alpha\varrho)ds,\qquad
    \varrho=\epsilon^{\mu\nu\lambda}x'_\mu x''_\nu x'''_\lambda,
\end{equation}
where $\alpha$ is a dimensionless parameter, and $\varrho$ is the
particle worldline trajectory torsion. Unlike the model
(\ref{Acurv}), here the parameter $\alpha$ can take positive or
negative values, and for the sake of definiteness, we assume that
$\alpha>0$. Action (\ref{Ator}) with $\alpha=1/2$ appeared
originally in the Euclidean version in the context of the Bose-Fermi
transmutation mechanism \cite{Pol2,MacWil}. Like the model of the
particle with curvature (\ref{Acurv}), the higher derivative system
(\ref{Ator}) possesses the  translation invariant dynamical spin
degrees of freedom $J_\mu=-\alpha e_\mu$,
$e_\mu=\dot{x}_\mu/\sqrt{-\dot{x}^2}$, as well as the three types of
solutions to the classical equations of motion, with $P^2<0$,
$P^2=0$ and $P^2>0$  \cite{Ptor}. At the quantum level operators
$J_\mu$ satisfy the SO(2,1) commutation relations
\begin{equation}\label{JJ}
    [J_\mu,J_\mu]=-i\epsilon_{\mu\nu\lambda}J^\lambda,
\end{equation}
analogous to those for the (2+1)D $\gamma$-matrices. Note that in
(2+1)D, there is a duality relation
$J_\mu=-\frac{1}{2}\epsilon_{\mu\nu\lambda}S^{\nu\lambda}$ between
the (2+1)D vector $J_\mu$ and the spin tensor $S_{\mu\nu}$
satisfying the commutation relations of the form (\ref{SS}). The
parameter $\alpha$ is not quantized here, and it fixes the value of
the Casimir operator of the algebra (\ref{JJ}),
$J^2=-\alpha(\alpha-1)$ \cite{Ptor}. For the gauge $\tau=x^0$, in
representation where the  operator $J_0$ is diagonal, its
eigenvalues are $j_0=\alpha+n$, $n=0,1,\ldots$. This means that the
spin degrees of freedom of the system realize a bounded from below
unitary infinite-dimensional representation $D^+_\alpha$ of the
universal covering group of the (2+1)D Lorentz group
\cite{Barg,SL2}. The physical states of the system are given by the
quantum analogue of the constraint responsible for the
reparametrization invariance of the action (\ref{Ator}) \cite{Ptor},
\begin{equation}\label{MP}
    (PJ-\alpha m)\Psi=0.
\end{equation}
One can treat Eq. (\ref{MP}) as  a (2+1)D analogue of the original
Majorana equation (\ref{Maj}). The difference of the (2+1)D  from
the (3+1)D case proceeds from the isomorphism between SO(2,2) and
SO(2,1)$\oplus$SO(2,1) algebras, and here the SO(2,1) generators
$J_\mu$ simultaneously play the role analogous to that played by the
SO(3,2) generators $\Gamma_\mu$  satisfying the commutation
relations (\ref{Gamma}). In the time-like sector,  the solutions of
Eq. (\ref{MP}) describe the positive energy states of the spin
$s_n=\alpha+n$ lying on the Majorana type trajectory \cite{Ptor}
\begin{equation}\label{Mn}
    M_n=\frac{m}{1+\alpha^{-1}n},\qquad n=0,1,\ldots.
\end{equation}

\section{Fractional spin  fields}

The (2+1)D analogue of the Majorana equation (\ref{MP}) being
supplied with the Klein-Gordon equation (\ref{KG}) describes the
fields carrying irreducible representation of the Poincar\'e
ISO(2,1) group of any, but fixed spin $s=\alpha>0$ \cite{Ptor}, and
so, can serve as a basis for relativistic  anyon theory
\cite{LeiMyr}--\cite{PlAny}. Instead of these two equations, one can
obtain the same result starting from the linear differential (2+1)D
Majorana-Dirac wave equations suggested in
\cite{PMajDir}\footnote{Jackiw and Nair \cite{JNany} proposed an
alternative theory based on the (2+1)D Majorana equation supplied
with the equation for topologically massive vector gauge field.}. In
such a case it is supposed that besides the index $n$ associated
with the infinite-dimensional half-bounded unitary representation
$D^+_\alpha$, the infinite-component field carries in addition a
spinor index, and that it satisfies Eq. (\ref{MP}) as well as the
Dirac equation
\begin{equation}\label{MD}
    (P\gamma-m)\Psi=0.
\end{equation}
As a consequence of Eqs. (\ref{MP}), (\ref{MD}), the Majorana-Dirac
field satisfies not only the Klein-Gordon equation, but also the
equations
\begin{equation}\label{JPg}
    (J\gamma+\alpha)\Psi=0,\qquad
    \epsilon_{\mu\nu\lambda}J^\mu\gamma^\nu P^\lambda\Psi=0,
\end{equation}
and one finds that it describes the positive energy states of the
mass $m$ and spin $s=\alpha-\frac{1}{2}$ \cite{PMajDir}.

The alternative way to describe an anyon field of the fixed mass and
spin consists in the construction of the (2+1)D analogue of the
Dirac spinor set of equations (\ref{Dirac}) generating the Majorana
and Klein-Gordon equations in the form of integrability conditions.
The construction needs the application of the so called deformed
Heisenberg algebra with reflection intimately  related to parabosons
\cite{Hdef,nSUSY},
\begin{equation}\label{Rdef}
    [a^-,a^+]=1+\nu R,\qquad
    R^2=1,\qquad
    \{a^\pm,R\}=0,
\end{equation}
where $\nu$ is a real deformation parameter. Here operator
$N=\frac{1}{2}\{a^+,a^-\}-\frac{1}{2}(\nu+1)$ plays the role of a
number operator, $[N,a^\pm]=\pm a^\pm$,  allowing us to present a
reflection operator $R$ in terms of $a^\pm$: $R=(-1)^N=\cos \pi N$.
For $\nu>-1$ algebra (\ref{Rdef}) admits infinite-dimensional
unitary representations realized on a Fock space\footnote{For
negative odd integer values $\nu=-(2k+1)$, $k=1,2\ldots$, the
algebra has finite, (2k+1)-dimensional nonunitary representations
\cite{Hdef}.}. In terms of operators $a^\pm$ the SO(2,1) generators
(\ref{JJ}) are realized in a quadratic form,
\begin{equation}\label{Jaa}
    J_0=\frac{1}{4}\{a^+,a^-\},\qquad
    J_\pm=J_1\pm iJ_2=\frac{1}{2}(a^\pm)^2.
\end{equation}
Here $J_\mu J^\mu=-s(s-1)$ with
$s=\frac{1}{4}(1\pm\nu)$ on the even/odd eigensubspaces of the
reflection operator $R$, i.e. as in the (3+1)D case
we have a direct sum of the two
infinite-dimensional irreducible representations of the (2+1)D
Lorentz group. These quadratic
operators together with linear operators
\begin{equation}\label{La}
    L_1=\frac{1}{\sqrt{2}}(a^++a^-),\qquad
    L_2=\frac{i}{\sqrt{2}}(a^+-a^-),
\end{equation}
extend the SO(2,1) algebra into the OSP(1$\vert$2) superalgebra:
\begin{equation}\label{LL}
    \{L_\alpha,L_\beta\}=4i(J\gamma)_{\alpha\beta},\qquad
    [J_\mu,L_\alpha]=\frac{1}{2}(\gamma_\mu L)_\alpha,
\end{equation}
where the (2+1)D $\gamma$-matrices are taken in the Majorana
representation,
$(\gamma_0)_\alpha{}^\beta=(\sigma_2)_\alpha{}^\beta$,
$(\gamma_1)_\alpha{}^\beta=i(\sigma_1)_\alpha{}^\beta$,
$(\gamma_2)_\alpha{}^\beta=i(\sigma_3)_\alpha{}^\beta$, and
$(\gamma_\mu)_{\alpha\beta}=(\gamma_\mu)_\alpha{}^\rho\epsilon_{\rho\beta}$.
With these ingredients, the (2+1)D analogue of the Dirac spinor set
of wave equations (\ref{Dirac}) is \cite{PanyDef}
\begin{equation}\label{Di2}
    \left( (P\gamma)_\alpha{}^\beta+m\epsilon_{\alpha}{}^\beta\right)L_\beta\Psi=0.
\end{equation}
From these two ($\alpha=1,2$) equations the (2+1)D Majorana and
Klein-Gordon equations appear in the form of integrability
conditions.

The spinor set of equations (\ref{Di2}) was used, in particular, for
investigation of the Lorentz symmetry breaking in the (3+1)D
massless theories with fractional helicity states \cite{KPT}.

\section{Exotic Galilei group and noncommutative plane}

A special non-relativistic limit ($c$ is a speed of light)
\cite{JNlim,DHNR}
\begin{equation}\label{JN}
    c\rightarrow \infty,\qquad
    s\rightarrow \infty,\qquad
    \frac{s}{c^2}=\kappa,
\end{equation}
applied to the spinor set of equations (\ref{Di2}) results in the
infinite-component Dirac-Majorana-L\'evy-Leblond type wave equations
\cite{HP2}
\begin{equation}\label{HP1}
    i\partial_t\phi_k+\sqrt{\frac{k+1}{2\theta}}\,\frac{P_+}{m}\,\phi_{k+1}=0,
\end{equation}
\begin{equation}\label{HP2}
    P_-\phi_k+\sqrt{\frac{2(k+1)}{\theta}}\,\phi_{k+1}=0,
\end{equation}
where $k=0,1,\ldots$, $P_\pm=P_1\pm iP_2$, and
\begin{equation}\label{the}
    \theta=\frac{\kappa}{m^2}.
\end{equation}
The first equation (\ref{HP1}) defines the
dynamics. The second equation relates different components of the
field allowing us to present them in terms of the lowest component,
\begin{equation}\label{phik}
    \phi_k=(-1)^k\left(\frac{\kappa}{2}\right)^{\frac{k}{2}}
    \left(\frac{P_-}{m}\right)^k\phi_0.
\end{equation}
Though a simple substitution of the second equation into the first
one shows that every component $\phi_k$ satisfies the Shr\"odinger
equation of a free planar particle, the nontrivial nature of the
system is encoded in its symmetry. The (2+1)D Poincar\'e symmetry of
the original relativistic system in the limit (\ref{JN}) is
transformed into the exotic planar Galilei symmetry characterized by
the noncommutative boosts \cite{LL,DH},
\begin{equation}\label{KK}
    [{\cal K}_1,{\cal K}_2]=-i\kappa.
\end{equation}
The system of  the two infinite-component equations (\ref{HP1}),
(\ref{HP2}) can be presented in the equivalent form
\begin{equation}\label{HV-}
    i\partial_t\phi=H\phi,\qquad V_-\phi=0,
\end{equation}
with
\begin{equation}\label{Hv}
        H=P_iv_i-\frac{1}{2}mv_+v_-\, ,\qquad
         V_-=v_--\frac{P_-}{m}\, .
\end{equation}
The translation invariant operators $v_\pm=v_1\pm iv_2$,
$[v_i,v_j]=-i\kappa^{-1}\epsilon_{ij}$, is the  non-relativistic
limit (\ref{JN}) of the noncompact Lorentz generators,
$-(c/s)J_\pm\rightarrow v_\pm$. The symmetry of the quantum
mechanical system (\ref{HV-}) is given by the  Hamiltonian $H$, the
space translation generators $P_i$, and by the rotation and boost
generators,
\begin{equation}\label{JK}
    {\cal J}=\epsilon_{ij}x_iP_j+\frac{1}{2}\kappa v_+v_-,\qquad
    {\cal K}_i=mx_i-tP_i+\kappa\epsilon_{ij}v_j.
\end{equation}
These integrals generate the algebra of the two-fold centrally
extended planar Galilei group \cite{LL,DH} characterized by the
non-commutativity of the boosts (\ref{KK}).

The first equation from (\ref{HV-}) is nothing  else as
a non-relativistic limit of the (2+1)D Majorana equation (\ref{MP})
\cite{HP2}.
The system described by it (without the second equation from
(\ref{HV-}))  corresponds to the classical system given by the
higher derivative Lagrangian
\begin{equation}\label{Ltheta}
    L=\frac{1}{2}m\dot{x}_i\,{}^2
    +\kappa\epsilon_{ij}\dot{x}_i\ddot{x}_j,
\end{equation}
 which, in its turn, corresponds to the non-relativistic limit (\ref{JN}) of the
 relativistic model of the particle with torsion (\ref{Ator})
 \cite{HP1}.
 It is interesting to note that the  system (\ref{Ltheta})
 (for the first time  considered by Lukierski, Stichel and Zakrzewski
  \cite{LSZ}, in  ignorance
 of its relation to the relativistic higher derivative model
 (\ref{Ator})), reveals the same
 dynamics as a charged non-relativistic planar particle in external
 homogeneous magnetic and electric fields \cite{OlP}.
The spectrum of the Hamiltonian (\ref{Hv}),
\begin{equation}\label{En}
    E_n(P)=\frac{1}{2m}P_i^2-m\kappa^{-1}n,\qquad
     n=0,1,\ldots,
\end{equation}
 is not
restricted from below, and the system (\ref{Ltheta}), similarly to
its relativistic analogue (\ref{Ator}), describes a reducible
representation of the exotic Galilei group. The role of the second
equation from (\ref{HV-}), whose component form is given by Eq.
(\ref{HP2}), consists in singling out the highest (at fixed $P_i^2$)
energy state from (\ref{En}) with $n=0$, and fixing an irreducible
infinite-dimensional unitary representation of the exotic planar
Galilei group \cite{HP2,OlP}. The system being reduced to the
surface given by this  second equation (classically equivalent to
the set of the two second class constraints $V_i=0$, $i=1,2$)
corresponds to the exotic planar particle considered by Duval and
Horvathy \cite{DH,Horv06}, which is described by the free particle
Hamiltonian and an exotic symplectic two-form,
\begin{equation}\label{DH}
    H=\frac{1}{2m}P_i^2\, ,\qquad \omega=dP_i\wedge dx_i +
    \frac{1}{2}\theta\epsilon_{ij}dP_i\wedge
    dP_j.
\end{equation}
The system (\ref{DH}) reveals a noncommutative structure encoded in
the nontrivial commutation relations of the particle coordinates,
\begin{equation}\label{xxt}
    [x_i,x_j]=i\theta\epsilon_{ij}.
\end{equation}
This noncommutative structure is the non-relativistic limit
(\ref{JN}) \cite{JNlim}  of the commutation relations
\begin{equation}\label{XXmin}
    [x_\mu,x_\nu]=-is\epsilon_{\mu\nu\lambda}\frac{P^\lambda}{(-P^2)^{3/2}}
\end{equation}
 associated with the minimal canonical approach for relativistic
 anyon of spin $s$ \cite{CorP}. Note that as was observed by Schonfeld
 \cite{Schon} (see also \cite{PCMon}),
 the commutation relations (\ref{XXmin}) are dual to the (Euclidean) commutation
 relations for the mechanical momentum  of a charged particle in the magnetic monopole
 field. The latter system also admits a description by the higher
 derivative Lagrangian \cite{PCMon},
\begin{equation}\label{Lmono}
    L_{CM}=\frac{1}{2}m\dot{\vec{r}}\,{}^2-eg
    \frac{\vert\vec{r}\vert}{(\vec{r}\times\dot{\vec{r}})^2}
    (\vec{r}\times\dot{\vec{r}})\cdot \ddot{\vec{r}}\, .
\end{equation}
There is a close relationship between the charge-monopole
non-relativistic system (\ref{Lmono})  and the model of relativistic
particle with torsion (\ref{Ator}). Indeed, in a parametrization
$x_\mu=x_\mu(\tau)$, the torsion term from (\ref{Ator}) takes the
(Minkowski) form of the higher derivative charge-monopole coupling
term, but in the \emph{velocity} space with $v^\mu\equiv
\dot{x}^\mu$,
\begin{equation}\label{tormono}
    L_{tor}=-\alpha\frac{\sqrt{-v^2}}{(\epsilon_{\gamma\rho\sigma}v^\rho
    \dot{v}^\sigma)^2}\epsilon_{\mu\nu\lambda}v^\mu
    \dot{v}^\nu\ddot{v}^\lambda.
\end{equation}
For system (\ref{Lmono}) the relation $\vec{J}\vec{n}+eg=0$ is the
analogue of the (2+1)D Majorana equation (\ref{MP}), where
$\vec{n}=\vec{r}/\vert\vec{r}\vert$ and $\vec{J}$ is the
charge-monopole angular momentum.

The exotic planar particle described by the  symplectic structure
(\ref{DH}), or by the Dirac-Majorana-L\'evy-Leblond type equations
(\ref{HP1}), (\ref{HP2}),  can be consistently coupled to an
arbitrary external  electromagnetic field at the \emph{classical}
level \cite{DH,HP3}. However, at the quantum level the Hamiltonian
reveals a nonlocal structure in the case of inhomogeneous magnetic
field \cite{HP3}. Another peculiarity reveals even in the case of
homogeneous magnetic field corresponding to the Landau problem for a
particle in a noncommutative plane \cite{HP3,P06,BelNer}, where the
initial particle mass $m$ is changed for the effective mass
\cite{DH}
\begin{equation}\label{mef}
    m^*=m(1-eB\theta),
\end{equation}
see below. As a result, the system develops three essentially
different phases corresponding to the subcritical, $eB\theta<1$,
critical, $eB\theta=1$, and overcritical, $eB\theta>1$, values of
the magnetic field \cite{HP3,P06}.

\section{Gauge theories with Chern-Simons terms and
exotic particle}

In the case of the choice of finite-dimensional non-unitary
representations of the deformed Heisenberg algebra with reflection
(\ref{Rdef}) corresponding to the negative odd values of the
deformation parameter, $\nu=-(2k+1)$, $k=1,2,\ldots$, the (2+1)D
spinor set of equations (\ref{Di2}) describes a spin-$j$ field with
$j=k/2$ and both signs of the energy \cite{Psusy,Puniver,CPvec}. In
particular, in the simplest cases of $\nu=-3$ and $\nu=-5$,
Eq. (\ref{Di2}) gives rise, respectively, to the Dirac
spin-$1/2$ particle theory and to the topologically massive
electrodynamics \cite{Schon,DSTem}. The latter system is described
by the Lagrangian
\begin{equation}\label{TPE}
    L_{TME}=-\frac{1}{4}F_{\mu\nu}F^{\mu\nu}-\frac{m}{4}\epsilon^{\mu\nu\lambda}A_{\mu}
    F_{\nu\lambda},\qquad
    F_{\mu\nu}=\partial_\mu A_\nu-\partial_\nu A_\mu.
\end{equation}
Let us suppress the dependence on the spatial coordinates $x_i$ by
making a substitution $A^\mu(x)\rightarrow \sqrt{m}\,r^\mu(t)$. Then
(\ref{TPE}) takes a form of the Lagrangian of a non-relativistic
charged particle in the homogeneous magnetic field $B=m^2e^{-1}$, $
    L=\frac{1}{2}m \dot{r}_i^2+\frac{1}{2}eB\epsilon_{ij}r_i\dot{r}_j,
$ while the variable $r^0$ disappears\footnote{This corresponds to
the nature of the $A^0$ field, which can be removed by imposing the
Weyl gauge $A^0=0$.}.

In Ref. \cite{DesJack3}, Deser and Jackiw proposed an extension of
the topologically massive electrodynamics by adding to Lagrangian
(\ref{TPE}) the higher derivative term of the Chern-Simons form,
\begin{equation}\label{DJ}
    L_{DJ}=L_{TME}+L_{ECS},\qquad
    L_{ECS}=\kappa m^{-1}
    \epsilon^{\mu\nu\lambda}F_{\mu\sigma}\partial^\sigma F_{\nu\lambda},
\end{equation}
where  $\kappa$ is a dimensionless numerical parameter. Making the
same substitution as before, and changing $r_i\rightarrow x_i$, we
reduce the (2+1)D field Lagrangian (\ref{DJ}) to the mechanical
Lagrangian for a particle in a plane,
\begin{equation}\label{NCB}
    L=\frac{1}{2}m
    \dot{x}_i^2+\frac{1}{2}eB\epsilon_{ij}x_i\dot{x}_j+\kappa\epsilon_{ij}\dot{x}_i\ddot{x}_j,
\end{equation}
that describes the higher derivative model (\ref{Ltheta}) coupled to
the external homogeneous magnetic field. The system (\ref{NCB}),
like the free higher derivative system (\ref{Ltheta}) underlying the
special non-relativistic limit (\ref{JN}) of the (2+1)D Majorana
equation, has a spectrum unbounded from below. This drawback can be
removed by supplying the coupled system with the appropriately
modified constraint (\ref{HP2}) \cite{HP3}. Classically, this is
equivalent  to the change of the higher derivative Lagrangian
(\ref{NCB}) for the first order exotic Duval-Horvathy Lagrangian
\cite{DH}
\begin{equation}\label{HDL}
    L_{ex}=P_i\dot{x}_i-\frac{1}{2m}P_i^2+\frac{1}{2}
    \theta\epsilon_{ij}P_i\dot{P}_j+\frac{1}{2}eB\epsilon_{ij}x_i\dot{x}_j,
\end{equation}
corresponding in a free case to the symplectic structure
(\ref{DH})\footnote{For the system (\ref{NCB}), one can get rid of
the unbounded from below spectrum  by changing the sign in the
first, kinetic term. In this case the problem  reappears at
$\kappa=0$.}. It generates the equations of motion with the
effective mass (\ref{mef}), $P_i=m^*\dot{x}_i$,
$\dot{P}_i=eB\epsilon_{ij}\dot{x}_j$.

The interacting exotic particle system (\ref{HDL}) can also be
obtained by a reduction of another (2+1)-dimensional Abelian gauge
field theory given by the Lagrangian with several  Chern-Simons
terms,
\begin{equation}\label{Hag}
    L_H=-\epsilon^{\mu\nu\lambda}\Phi_\mu \partial_\nu A_\lambda
     -\frac{1}{2}\lambda\Phi_\mu\Phi^\mu-\frac{1}{2}\kappa
     m^{-1}\epsilon^{\mu\nu\lambda}
     \Phi_\mu\partial_\nu\Phi_\lambda-\frac{1}{2}\beta
     m\epsilon^{\mu\nu\lambda}A_\mu\partial_\nu A_\lambda,
\end{equation}
where $\lambda$, $\kappa$ and $\beta$ are dimensionless parameters.
The system with Lagrangian (\ref{Hag}) was investigated by Hagen
\cite{Hagen}, see also \cite{DesJack2}. Suppressing the dependence
of the fields $\Phi_\mu$ and $A_\mu$ on the spatial coordinates by
making the substitutions $A^\mu(x)\rightarrow \sqrt{\lambda m}\,
r^\mu(t)$ and $\Phi^\mu(x)\rightarrow \pi^\mu(t)/\sqrt{m\lambda}$
(we assume $\lambda>0$), and denoting $\lambda\beta m^2=eB$ and
$\kappa/(\lambda m^2)=\theta$, we reduce (\ref{Hag}) to the first
order Lagrangian
$$
    L=\epsilon_{ij}\pi_i\dot{r}_j
    -\frac{1}{2m}\pi_i^2+\frac{1}{2m}\pi_0^2+\frac{1}{2}
    \theta\epsilon_{ij}\pi_i\dot{\pi}_j+\frac{1}{2}eB\epsilon_{ij}r_i\dot{r}_j.
$$
Hence, the $\pi_0$ plays the role of the auxiliary variable, and can
be omitted using its equation of motion
$\pi_0=0$\footnote{Disappearance of $\pi_0$ ($r_0$) is rooted in the
independence of Lagrangian (\ref{Hag}) of the time derivative of
$\Phi^0$ ($A^0$).}. Then, changing the notations $r_i\rightarrow
x_i$ and $\pi_i\rightarrow \epsilon_{ij}P_j$, we arrive at the
Lagrangian (\ref{HDL}).

Therefore, the both systems (\ref{NCB}) and (\ref{HDL}),
corresponding  (in a free case) to the special non-relativistic
limit (\ref{JN}) of the (2+1)D Majorana equation (\ref{MP}) and
Dirac spinor set of equations (\ref{Di2}),  can be treated as
reduced versions of the relativistic  Lagrangians (\ref{DJ}) and
(\ref{Hag}) of  the (2+1)D Abelian gauge field theories with
Chern-Simons terms.

\section{Conclusion}

To conclude, we point out two interesting open problems related to
the Majorana equation.

 It is known that the spin-statistics connection
for the infinite-component fields described by the Majorana type
equations is absent \cite{AGN,StoTod,BoLoToOk}.  On the other hand,
the question on such a connection for the fields of fixed mass and
spin described by the Dirac covariant set of equations is open. The
question on the spin-statistics relation for the fractional spin
field theories constructed on the basis of the (2+1)D analogue of
the Majorana equation  also still waits for the solution.

As we saw, the original (3+1)D Majorana equation and the Dirac
spinor set of equations constructed on its basis, as well as their
(2+1)D analogues,  have a hidden supersymmetric structure encoded in
the Majorana spectrum (\ref{Mj}). Hence, it would be very natural to
try to construct a supersymmetric extension  of these theories. Such
an attempt was undertaken in Ref. \cite{Puniver} for the case of the
(2+1)D analogue of the Dirac spinor set of equations. Within the
framework of a restricted approach taken there, the supersymmetric
extension was obtained only for a few special cases corresponding to
finite-dimensional representations of the underlying deformed
Heisenberg algebra with reflection (\ref{Rdef})\footnote{See also
ref. \cite{DVS} for the case of $\nu=0$. }. A supersymmetric
extension could help to resolve the problem of the electromagnetic
coupling, including the quantum case of the non-relativistic exotic
particle in the noncommutative plane.

\vskip 0.1cm\noindent {\bf Acknowledgements}. The work was supported
in part by  FONDECYT grant 1050001.

%%%%%%%%%%%%%%%%%%%%%%%%%%%%%%%%%%%%%%%%%%%%%%%%%%%%%%%%%%%%
%%%%%%%%%%%%%%%%%%%%%%%%%%%%%%%%%%%%%%%%%%%%%%%%%%%%%%%%%%%%


\begin{thebibliography}{99}
%%%%%%%%%%%%%%%%%%%%%%%%%%%%%%%%%%%%%%%%%%%%%%%%%%%%%%%%%%%%
%%%%%%%%%%%%%%%%%%%%%%%%%%%%%%%%%%%%%%%%%%%%%%%%%%%%%%%%%%%%
\bibitem{Maj}
E. Majorana, \emph{Teoria relativistica di particelle con momento
intrinseco arbitrario}, \textsl{ Nuovo Cimento} {\bf 9} (1932) 335.

\bibitem{GelYa}
I. M. Gel'fand and A. M. Yaglom, \textsl{Zh. Eksp. Teor. Fiz.}
\textbf{18} (1948) 703, 1096, 1105.

\bibitem{Frad}
M. Fradkin, \textsl{Am. J. of Physics} {\bf 34} (1966) 314.

\bibitem{Fro1}
C. Fronsdal, \textsl{Phys. Rev.} \textbf{156} (1967) 1665;
  %``Infinite-Component Field Theories, Fubini Sum Rules Completeness, And
  %Current Algebra. I. Discrete Spectra,''
 \textsl{ Phys.\ Rev. } {\bf 182} (1969) 1564;
  %%CITATION = PHRVA,182,1564;%%
C. Fronsdal and R. White, \textsl{Phys. Rev.} \textbf{163} (1967)
1835.

\bibitem{AGN}
E. Abers, I. T. Grodsky and R. E. Norton,
% Diseases of infinite-component field theories
\textsl{Phys. Rev.} \textbf{159} (1967) 1222.

\bibitem{Nambu}
Y. Nambu, \emph{Phys. Rev.} \textbf{160} (1967) 1171.

\bibitem{StoTod}
D. Tz. Stoyanov and I. T. Todorov, \textsl{J. Math. Phys.}
\textbf{9} (1968) 2146.

\bibitem{BoLoToOk}
  N.~N.~Bogolyubov, A.~A.~Logunov, A.~I.~Oksak and I.~T.~Todorov,
  \emph{General Principles Of Quantum Field Theory}
    Dordrecht, Netherlands: Kluwer (1990).
%\href{http://www.slac.stanford.edu/spires/find/hep/www?irn=2277816}{SPIRES entry}

\bibitem{Venez}
  G.~Veneziano,
  %``Construction Of A Crossing - Symmetric, Regge Behaved Amplitude For
  %Linearly Rising Trajectories,''
 \textsl{ Nuovo Cim. A}  {\bf 57} (1968) 190.
  %%CITATION = NUCIA,A57,190;%%

\bibitem{NevSch}
  A.~Neveu and J.~H.~Schwarz,
  %``Factorizable Dual Model Of Pions,''
\textsl{  Nucl.\ Phys. B}  {\bf 31} (1971) 86.
  %%CITATION = NUPHA,B31,86;%%

\bibitem{Ramon}
  P.~Ramond,
  %``Dual Theory For Free Fermions,''
 \textsl{ Phys.\ Rev. D}  {\bf 3}, 2415 (1971).
  %%CITATION = PHRVA,D3,2415;%%

\bibitem{Mandelstam}
  S.~Mandelstam,
  %``Dual - Resonance Models,''
 \textsl{ Phys.\ Rept. } {\bf 13} (1974) 259.
  %%CITATION = PRPLC,13,259;%%

\bibitem{Gliozzi:1976jf}
  F.~Gliozzi, J.~Scherk and D.~I.~Olive,
  %``Supergravity And The Spinor Dual Model,''
 \textsl{ Phys.\ Lett. B}  {\bf 65} (1976) 282.
  %%CITATION = PHLTA,B65,282;%%

\bibitem{Green:1980zg}
  M.~B.~Green and J.~H.~Schwarz,
  %``Supersymmetrical Dual String Theory,''
 \textsl{ Nucl.\ Phys. B}  {\bf 181} (1981) 502.
  %%CITATION = NUPHA,B181,502;%%

\bibitem{GSW}
  M.~B.~Green, J.~H.~Schwarz and E.~Witten,
 \emph{ Superstring Theory}, Cambridge University Press, Cambridge, 1987.

\bibitem{Dirac}
P. A. M. Dirac, \textsl{Proc. R. Soc. Lond. Ser. A} \textbf{322}
(1971) 435, and ibid. \textbf{328} (1972) 1.


\bibitem{Stau1}
  L.~P.~Staunton,
  %``Covariant Heisenberg Picture Of A Relativistic Positive - Energy Theory:
  %The Operator Algebra Of The Rigid String,''
 \textsl{ Phys.\ Rev. D}  {\bf 13} (1976) 3269;
  %%CITATION = PHRVA,D13,3269;%%
  L.~P.~Staunton and S.~Browne,
  %``The Classical Limit Of Relativistic Positive Energy Theories With Intrinsic
  %Spin,''
 \textsl{ Phys.\ Rev. D}  {\bf 12} (1975) 1026.
  %%CITATION = PHRVA,D12,1026;%%

\bibitem{Biedenharn}
  L.~C.~Biedenharn and H.~Van Dam,
  %``Galilean Subdynamics And The Dual Resonance Model,''
\textsl{  Phys.\ Rev. D}  {\bf 9} (1974) 471;
  %%CITATION = PHRVA,D9,471;%%
  %``The Kinematics Of A Poincare Covariant Object Having Indecomposable
  %Internal Structure,''
  {\bf 14} (1976) 405.
  %%CITATION = PHRVA,D14,405;%%

\bibitem{Barut}
  A.~O.~Barut and I.~H.~Duru,
  %``Relativistic Composite Systems And Minimal Coupling,''
\textsl{  Phys.\ Rev. D}  {\bf 10} (1974) 3448.
  %%CITATION = PHRVA,D10,3448;%%

\bibitem{Stau2}
  L.~P.~Staunton,
  %``A Spin 1/2 Positive Energy Relativistic Wave Equation,''
\textsl{  Phys.\ Rev. D}  {\bf 10} (1974) 1760.
  %%CITATION = PHRVA,D10,1760;%%

\bibitem{LeutS}
  H.~Leutwyler and J.~Stern,
  %``Relativistic Dynamics On A Null Plane,''
\textsl{  Annals Phys. }  {\bf 112} (1978) 94.
  %%CITATION = APNYA,112,94;%%


\bibitem{MukDam}
  N.~Mukunda, H.~Van Dam and L.~C.~Biedenharn,
  %``Composite Systems As Relativistic Quantal Rotators: Vectorial And Spinorial
  %Models,''
\textsl{  Phys.\ Rev. D}  {\bf 22} (1980) 1938.
  %%CITATION = PHRVA,D22,1938;%%


\bibitem{HP1}
P.~A.~Horv\'athy and M.~S.~Plyushchay,
%{\it Non-relativistic
%anyons, exotic Galilean symmetry and noncommutative plane}.
\textsl{JHEP} {\bf 0206} (2002) 033 [\texttt{hep-th/0201228}].
%%CITATION = HEP-TH 0201228;%%

\bibitem{HP2}
  P.~A.~Horvathy and M.~S.~Plyushchay,
  %``Anyon wave equations and the noncommutative plane,''
\textsl{  Phys.\ Lett. B}  {\bf 595} (2004) 547
  [\texttt{hep-th/0404137}].
  %%CITATION = HEP-TH 0404137;%%

\bibitem{HP3}
  P.~A.~Horvathy and M.~S.~Plyushchay,
  %``Nonrelativistic anyons in external electromagnetic field,''
   \textsl{ Nucl. Phys. B}  {\bf  714} (2005) 269
  [\texttt{hep-th/0502040}].
  %%CITATION = HEP-TH 0502040;%%

\bibitem{LeiMyr}
  J.~M.~Leinaas and J.~Myrheim,
  %``On The Theory Of Identical Particles,''
 \textsl{ Nuovo Cim. B}  {\bf 37} (1977) 1.
  %%CITATION = NUCIA,B37,1;%%

\bibitem{WilZ}
  F.~Wilczek and A.~Zee,
  %``Linking Numbers, Spin, And Statistics Of Solitons,''
 \textsl{ Phys.\ Rev.\ Lett. } {\bf 51} (1983) 2250.
  %%CITATION = PRLTA,51,2250;%%

\bibitem{WuZ}
  Y.~S.~Wu and A.~Zee,
  %``Comments On The Hopf Lagrangian And Fractional Statistics Of Solitons,''
\textsl{  Phys.\ Lett.\ B}  {\bf 147} (1984) 325.
  %%CITATION = PHLTA,B147,325;%%

\bibitem{MacWil}
R. MacKenzie, F. Wilczek,
%Peculiar Spin And Statistics In Two Space Dimensions.
%DOE/ER/01545-406, (Received Aug 1988). 52pp. Published in
\textsl{Int. J. Mod. Phys. A} {\bf 3} (1988) 2827.
 %%CITATION = IMPAE,A3,2827;%%

\bibitem{DV}
 D.~V.~Volkov,
  %``Quartions In Relativistic Field Theory,''
\textsl{  JETP Lett. } {\bf 49} (1989) 541.
%  [Pisma Zh.\ Eksp.\ Teor.\ Fiz.\  {\bf 49} (1989) 473].
  %%CITATION = JTPLA,49,541;%%


\bibitem{PlAny}
  M.~S.~Plyushchay,
  %``Relativistic Model Of the Anyon,''
 \textsl{ Phys.\ Lett. B}  {\bf 248} (1990) 107.
  %%CITATION = PHLTA,B248,107;%%

\bibitem{Ptor}
  M.~S.~Plyushchay,
  %``Relativistic Particle With Torsion, Majorana Equation And Fractional
  %Spin,''
 \textsl{ Phys.\ Lett. B}  {\bf 262} (1991) 71;
  %%CITATION = PHLTA,B262,71;%%
  %``The Model Of Relativistic Particle With Torsion,''
\textsl{  Nucl.\ Phys.\ B} {\bf 362} (1991) 54.
  %%CITATION = NUPHA,B362,54;%%

\bibitem{JNany}
    R. Jackiw and V. P. Nair,
    %{\it Relativistic wave equation for
    %anyons}.
    {\sl Phys. Rev. D} {\bf 43} (1991) 1933.
%%CITATION = PHRVA,D43,1933;%%

\bibitem{PMajDir}
  M.~S.~Plyushchay,
    %``Fractional spin: Majorana-Dirac field,''
\textsl{  Phys.\ Lett. B}  {\bf 273} (1991) 250;
  %%CITATION = PHLTA,B273,250;%%
  %``The model of a free relativistic particle with fractional spin,''
 \textsl{ Int.\ J.\ Mod.\ Phys.\ A }{\bf 7} (1992) 7045.
  %%CITATION = IMPAE,A7,7045;%%

\bibitem{DVS}
  D.~P.~Sorokin and D.~V.~Volkov,
  %``(Anti)commuting spinors and supersymmetric dynamics of semions,''
 \textsl{ Nucl.\ Phys.\ B} {\bf 409} (1993) 547.
  %%CITATION = NUPHA,B409,547;%%

\bibitem{CNPo}
  Ch.~Chou, V.~P.~Nair and A.~P.~Polychronakos,
  %``On the electromagnetic interactions of anyons,''
 \textsl{ Phys.\ Lett.\ B }{\bf 304} (1993) 105
  [\texttt{hep-th/9301037}].
  %%CITATION = HEP-TH 9301037;%%


\bibitem{CPvec}
  J.~L.~Cortes and M.~S.~Plyushchay,
  %``Linear differential equations for a fractional spin field,''
\textsl{  J.\ Math.\ Phys. } {\bf 35} (1994) 6049
  [\texttt{hep-th/9405193}].
  %%CITATION = HEP-TH 9405193;%%

\bibitem{PanyDef}
 M.~S.~Plyushchay,
  %``Deformed Heisenberg algebra and fractional spin field in
  %(2+1)-dimensions,''
    {\sl Phys. Lett. B}  {\bf  320} (1994) 91
    [\texttt{hep-th/9309148}] .

\bibitem{LL}
J.-M.~L\'evy-Leblond, {\it Galilei group and Galilean invariance}.
In: {\it Group Theory and Applications} (Loebl Ed.), {\bf II}, Acad.
Press, New York, p. 222 (1972); A. Ballesteros, M. Gadella and
M.~del Olmo,
% {\it Moyal quantization of 2+1 dimensional Galilean
%systems}.
{\sl Journ. Math. Phys.} {\bf 33} (1992) 3379; Y.~Brihaye,
C.~Gonera, S.~Giller and P.~Kosi\'nski,
%{\it Galilean invariance in
%$2+1$ dimensions.}
\texttt{hep-th/9503046}.
%%CITATION = HEP-TH 9503046;%%

\bibitem{LSZ}
J.~Lukierski, P.~C.~Stichel, W.~J.~Zakrzewski,
    %{\it Galilean-invariant $(2+1)$-dimensional models with
    %a Chern-Simons-like term and $d=2$ noncommutative
    %geometry}.
    {\sl Annals of Physics (NY)} {\bf 260} (1997) 224
    [\texttt{hep-th/9612017}].
    %%CITATION = HEP-TH 9612017;%%

\bibitem{DH}
C.~ Duval and P.~A.~Horv\'athy,
    %{\it The exotic Galilei group and the ``Peierls
    %substitution''}.
{\sl Phys. Lett. B} {\bf  479} (2000) 284 [\texttt{hep-th/0002233}];
 %%CITATION = HEP-TH 0002233;%%
%{\it Exotic Galilean symmetry in the
%noncommutative plane, and the Hall  effect}.
{\sl Journ. Phys. A} {\bf  34} (2001) 10097
[\texttt{hep-th/0106089}]
%%CITATION = HEP-TH 0106089;%%

\bibitem{HorMar}
  P.~A.~Horvathy, L.~Martina and P.~C.~Stichel,
  %``Galilean symmetry in noncommutative field theory,''
{\sl  Phys.\ Lett.\ B} {\bf 564} (2003) 149
  [arXiv:hep-th/0304215].
  %%CITATION = HEP-TH 0304215;%%


\bibitem{SeiWit}
  N.~Seiberg and E.~Witten,
  %``String theory and noncommutative geometry,''
  JHEP {\bf 9909} (1999) 032
  [\texttt{hep-th/9908142}].
  %%CITATION = HEP-TH 9908142;%%


\bibitem{BigSus}
  D.~Bigatti and L.~Susskind,
  %``Magnetic fields, branes and noncommutative geometry,''
\textsl{  Phys.\ Rev.\ D} {\bf 62} (2000) 066004
  [\texttt{hep-th/9908056}].
  %%CITATION = HEP-TH 9908056;%%

\bibitem{DougNek}
  M.~R.~Douglas and N.~A.~Nekrasov,
  %``Noncommutative field theory,''
\textsl{  Rev.\ Mod.\ Phys. } {\bf 73} (2001) 977
  [\texttt{hep-th/0106048}].
  %%CITATION = HEP-TH 0106048;%%


\bibitem{ConS}
  A.~Connes, M.~R.~Douglas and A.~S.~Schwarz,
  %``Noncommutative geometry and matrix theory: Compactification on tori,''
 \textsl{ JHEP} {\bf 9802} (1998) 003
  [\texttt{hep-th/9711162}].
  %%CITATION = HEP-TH 9711162;%%


\bibitem{BerMar}
  F.~A.~Berezin and M.~S.~Marinov,
  %``Particle Spin Dynamics As The Grassmann Variant Of Classical Mechanics,''
 \textsl{ Annals Phys. } {\bf 104} (1977) 336.
  %%CITATION = APNYA,104,336;%%

\bibitem{Pcurv1}
  M.~S.~Plyushchay,
  %``Canonical Quantization And Mass Spectrum Of Relativistic Particle: Analog
  %Of Relativistic String With Rigidity,''
 \textsl{ Mod.\ Phys.\ Lett.\ A} {\bf 3} (1988) 1299;
  %%CITATION = MPLAE,A3,1299;%%
  %``Massive Relativistic Point Particle With Rigidity,''
  \textsl{Int.\ J.\ Mod.\ Phys.\ A} {\bf 4} (1989) 3851.
  %%CITATION = IMPAE,A4,3851;%%

\bibitem{Pol1}
  A.~M.~Polyakov,
  %``Fine Structure Of Strings,''
 \textsl{ Nucl.\ Phys.\ B }{\bf 268} (1986) 406.
  %%CITATION = NUPHA,B268,406;%%

\bibitem{Pol2}
  A.~M.~Polyakov,
  %``Fermi-Bose Transmutations Induced By Gauge Fields,''
 \textsl{ Mod.\ Phys.\ Lett.\ A} {\bf 3} (1988) 325.
  %%CITATION = MPLAE,A3,325;%%


\bibitem{JNlim}
  R.~Jackiw and V.~P.~Nair,
    %  \textit{Anyon spin and the exotic central extension of the planar Galilei
    % group}.
    \textsl{Phys. Lett. B}  {\bf  480} (2000) 237
    [\texttt{hep-th/0003130}].
  %%CITATION = HEP-TH 0003130;%%

\bibitem{DHNR}
  C.~Duval and P.~A.~Horvathy,
  %``Spin and exotic Galilean symmetry,''
  Phys.\ Lett.\ B {\bf 547} (2002) 306
  [arXiv:hep-th/0209166].
  %%CITATION = HEP-TH 0209166;%%


\bibitem{HorLan}
  P.~A.~Horvathy,
  %``The non-commutative Landau problem, and the Peierls substitution,''
  \textsl{Annals Phys. } {\bf 299} (2002) 128
  [\texttt{hep-th/0201007}].
  %%CITATION = HEP-TH 0201007;%%



\bibitem{P06}
  M.~S.~Plyushchay,
  %``Anyons and the Landau problem in the noncommutative plane,''
  \texttt{hep-th/0603034}.
  %%CITATION = HEP-TH 0603034;%%

\bibitem{Schon}
  J.~F.~Schonfeld,
  %``A Mass Term For Three-Dimensional Gauge Fields,''
 \textsl{ Nucl.\ Phys.\ B} {\bf 185} (1981) 157.
  %%CITATION = NUPHA,B185,157;%%


\bibitem{DSTem}
  S.~Deser, R.~Jackiw and S.~Templeton,
  %``Three-Dimensional Massive Gauge Theories,''
  \textsl{Phys.\ Rev.\ Lett. } {\bf 48} (1982) 975;
  %%CITATION = PRLTA,48,975;%%
  %``Topologically Massive Gauge Theories,''
\textsl{  Annals Phys. } {\bf 140} (1982) 372.
  %%CITATION = APNYA,140,372;%%


\bibitem{DesJack1}
  S.~Deser and R.~Jackiw,
  %``'Selfduality' Of Topologically Massive Gauge Theories,''
 \textsl{ Phys.\ Lett.\ B} {\bf 139} (1984) 371.
  %%CITATION = PHLTA,B139,371;%%


\bibitem{DinZ}
  A.~M.~Din and W.~J.~Zakrzewski,
  %``Spin And Statistics Of Cp**1 Skyrmions,''
 \textsl{ Phys.\ Lett.\ B} {\bf 146} (1984) 341.
  %%CITATION = PHLTA,B146,341;%%

\bibitem{Hagen}
  C.~R.~Hagen,
  %``What Is The Most General Abelian Gauge Theory In Two Spatial Dimensions?,''
\textsl{  Phys.\ Rev.\ Lett.}  {\bf 58} (1987) 1074.
  %%CITATION = PRLTA,58,1074;%%


\bibitem{DesJack2}
  S.~Deser and R.~Jackiw,
  %``Comment On 'What Is The Most General Abelian Gauge Theory In Two Spatial
  %Dimensions?',''
\textsl{  Phys.\ Rev.\ Lett. } {\bf 59} (1987) 1981.
  %%CITATION = PRLTA,59,1981;%%

\bibitem{WitJon}
  E.~Witten,
  %``Quantum Field Theory And The Jones Polynomial,''
  \textsl{Commun.\ Math.\ Phys.}  {\bf 121} (1989) 351.
  %%CITATION = CMPHA,121,351;%%


\bibitem{DesJack3}
  S.~Deser and R.~Jackiw,
  %``Higher derivative Chern-Simons extensions,''
\textsl{  Phys.\ Lett.\ B} {\bf 451} (1999) 73
  [\texttt{hep-th/9901125}].
  %%CITATION = HEP-TH 9901125;%%


\bibitem{Pisar}
  R.~D.~Pisarski,
  %``Theory Of Curved Paths,''
  \textsl{Phys.\ Rev.\ D} {\bf 34} (1986) 670.
  %%CITATION = PHRVA,D34,670;%%

\bibitem{Nester}
  V.~V.~Nesterenko,
  %``The Singular Lagrangians With Higher Derivatives,''
\textsl{  J.\ Phys.\ A} {\bf 22} (1989) 1673.
  %%CITATION = JPAGB,A22,1673;%%



\bibitem{PKuz}
  Y.~A.~Kuznetsov and M.~S.~Plyushchay,
  %``The Model of the relativistic particle with curvature and torsion,''
 \textsl{ Nucl.\ Phys.\ B }{\bf 389} (1993) 181.
  %%CITATION = NUPHA,B389,181;%%


\bibitem{Pmas0}
 M.~S.~Plyushchay,
  %``Massless Point Particle With Rigidity,''
  \textsl{Mod.\ Phys.\ Lett.\ A} {\bf 4} (1989) 837;
  %%CITATION = MPLAE,A4,837;%%
  %``Massless Particle With Rigidity As A Model For Description Of Bosons And
  %Fermions,''
  \textsl{Phys.\ Lett.\ B} {\bf 243} (1990) 383.
  %%CITATION = PHLTA,B243,383;%%

\bibitem{Ramos}
  E.~Ramos and J.~Roca,
  %``W symmetry and the rigid particle,''
 \textsl{ Nucl.\ Phys.\ B} {\bf 436} (1995) 529
  [\texttt{hep-th/9408019}];
  %%CITATION = HEP-TH 9408019;%%
  %``Extended gauge invariance in geometrical particle models and the geometry
  %of W symmetry,''
   {\bf 452} (1995) 705
  [\texttt{hep-th/9504071}].
  %%CITATION = HEP-TH 9504071;%%


\bibitem{Mourad}
  J.~Mourad,
  %``Continuous spin and tensionless strings,''
  \texttt{hep-th/0410009}.
  %%CITATION = HEP-TH 0410009;%%

\bibitem{EdMar}
  L.~Edgren, R.~Marnelius and P.~Salomonson,
  %``Infinite spin particles,''
\textsl{  JHEP} {\bf 0505} (2005) 002
  [\texttt{hep-th/0503136}];
  %%CITATION = HEP-TH 0503136;%%
  L.~Edgren and R.~Marnelius,
  %``Covariant quantization of infinite spin particle models, and higher order
  %gauge theories,''
  \texttt{hep-th/0602088}.
  %%CITATION = HEP-TH 0602088;%%

\bibitem{HSF}
 C.~Fronsdal,
  %``Massless Fields With Integer Spin,''
 \textsl{ Phys.\ Rev.\ D} {\bf 18} (1978) 3624.
  %%CITATION = PHRVA,D18,3624;%%

\bibitem{HSV}
 M.~A.~Vasiliev,
  %``Nonlinear equations for symmetric massless higher spin fields in
  %(A)dS(d),''
\textsl{  Phys.\ Lett.\ B} {\bf 567} (2003) 139
  [\texttt{hep-th/0304049}];
  %%CITATION = HEP-TH 0304049;%%
  %``Higher spin gauge theories in any dimension,''
 \textsl{ Comptes Rendus Physique }{\bf 5} (2004) 1101
  [\texttt{hep-th/0409260}].
  %%CITATION = HEP-TH 0409260;%%



\bibitem{Barg}
  V.~Bargmann,
  %``Irreducible Unitary Representations Of The Lorentz Group,''
 \textsl{ Annals Math. } {\bf 48} (1947) 568.
  %%CITATION = ANMAA,48,568;%%

\bibitem{SL2}
  M.~S.~Plyushchay,
  %``Quantization of the classical SL(2,R) system and representations of SL(2,R)
  %group,''
\textsl{  J.\ Math.\ Phys.}  {\bf 34} (1993) 3954.
  %%CITATION = JMAPA,34,3954;%%



\bibitem{Hdef}
M. S. Plyushchay,
%%CITATION = HEP-TH 9601116;%%
%{\it Deformed Heisenberg algebra with reflection}.
{\sl Nucl. Phys. B} {\bf 491} [PM] (1997) 619
[\texttt{hep-th/9701091}];
%%CITATION = HEP-TH 9701091;%%

\bibitem{nSUSY}
  M.~Plyushchay,
  %``Supersymmetries in pure parabosonic systems,''
 \textsl{ Int.\ J.\ Mod.\ Phys.\ A} {\bf 15} (2000) 3679
  [\texttt{hep-th/9903130}].
  %%CITATION = HEP-TH 9903130;%%


\bibitem{KPT}
  S.~M.~Klishevich, M.~S.~Plyushchay and M.~Rausch de Traubenberg,
  %``Fractional helicity, Lorentz symmetry breaking and anyons,''
 \textsl{ Nucl.\ Phys.\ B} {\bf 616} (2001) 419
  [\texttt{hep-th/0101190}].
  %%CITATION = HEP-TH 0101190;%%

\bibitem{OlP}
  M.~A.~del Olmo and M.~S.~Plyushchay,
  %``Centaurus: Enlarged exotic Galilei symmetry, electric-translation
  %transmutation and noncommutative plane,''
 \textsl{ Ann. Phys. (NY)} (2006), in press
  [\texttt{hep-th/0508020}].
  %%CITATION = HEP-TH 0508020;%%


\bibitem{CorP}
  J.~L.~Cortes and M.~S.~Plyushchay,
  %``Anyons as spinning particles,''
  \textsl{Int.\ J.\ Mod.\ Phys.\ A} {\bf 11} (1996) 3331
  [\texttt{hep-th/9505117}].
  %%CITATION = HEP-TH 9505117;%%

\bibitem{Horv06}
  P.~A.~Horvathy,
  %``Exotic galilean symmetry and non-commutative mechanics in mathematical and
  %in condensed matter physics,''
 \texttt{ hep-th/0602133}.
  %%CITATION = HEP-TH 0602133;%%



\bibitem{PCMon}
  M.~S.~Plyushchay,
  %``Monopole Chern-Simons term: Charge-monopole system as a particle with
  %spin,''
 \textsl{ Nucl.\ Phys.\ B} {\bf 589} (2000) 413
  [\texttt{hep-th/0004032}].
  %%CITATION = HEP-TH 0004032;%%

\bibitem{BelNer}
  S.~Bellucci, A.~Nersessian and C.~Sochichiu,
  %``Two phases of the non-commutative quantum mechanics,''
  \textsl{Phys.\ Lett.\ B} {\bf 522} (2001) 345
  [\texttt{hep-th/0106138}].
  %%CITATION = HEP-TH 0106138;%%

\bibitem{Psusy}
M. S. Plyushchay,
%{\it Deformed Heisenberg algebra, fractional spin
%fields and supersymmetry without fermions}.
{\sl Ann. Phys. (NY)}  {\bf 245} (1996) 339
[\texttt{hep-th/9601116}].


\bibitem{Puniver}
  M.~S.~Plyushchay,
  %``R-deformed Heisenberg algebra, anyons and d = 2+1 supersymmetry,''
 \textsl{ Mod.\ Phys.\ Lett. A}  {\bf 12} (1997) 1153
  [\texttt{hep-th/9705034}].
  %%CITATION = HEP-TH 9705034;%%






\end{thebibliography}
\end{document}